# Towards Adversarial Realism and Robust Learning for IoT Intrusion Detection and Classification

João Vitorino[0000-0002-4968-3653], Isabel Praça[0000-0002-2519-9859] and Eva Maia[0000-0002-8075-531X]

Research Group on Intelligent Engineering and Computing for Advanced Innovation and Development (GECAD), School of Engineering, Polytechnic of Porto (ISEP/IPP), 4249-015 Porto, Portugal
`{jpmvo,icp,egm}@isep.ipp.pt`

**Abstract.** The Internet of Things (IoT) faces tremendous security challenges. Machine learning models can be used to tackle the growing number of cyber-attack variations targeting IoT systems, but the increasing threat posed by adversarial attacks restates the need for reliable defense strategies. This work describes the types of constraints required for a realistic adversarial cyber-attack example and proposes a methodology for a trustworthy adversarial robustness analysis with a realistic adversarial evasion attack vector. The proposed methodology was used to evaluate three supervised algorithms, Random Forest (RF), Extreme Gradient Boosting (XGB), and Light Gradient Boosting Machine (LGBM), and one unsupervised algorithm, Isolation Forest (IFOR). Constrained adversarial examples were generated with the Adaptative Perturbation Pattern Method (A2PM), and evasion attacks were performed against models created with regular and adversarial training. Even though RF was the least affected in binary classification, XGB consistently achieved the highest accuracy in multi-class classification. The obtained results evidence the inherent susceptibility of tree-based algorithms and ensembles to adversarial evasion attacks and demonstrates the benefits of adversarial training and a security by design approach for a more robust IoT network intrusion detection and cyber-attack classification.

## 1 Introduction

The Internet of Things (IoT) is accelerating the digital transformation. It represents decentralized and heterogeneous systems of interconnected devices, which combine wireless sensor networks, real-time computing, and actuation technologies [1]. Due to the integration of physical and business processes, as well as control and information systems, IoT is bridging the gap between operational technology and information technology [2]. However, the convergence of previously isolated systems and technologies faces tremendous security challenges because of the software vulnerabilities and weak security measures of IoT devices [3]. A self-propagating malware can compromise numerous devices and establish a botnet to launch a wide range of cyber-attacks [4], which is particularly concerning for IoT systems that control critical infrastructure like healthcare facilities [5], energy markets [6], and water supply networks [7].

Machine Learning (ML) can be very valuable to tackle the growing number and increasing sophistication of cyber-attacks targeting IoT systems, but it is susceptible to adversarial examples: cyber-attack variations specifically crafted to exploit ML [8]. For instance, tree-based algorithms and ensembles are remarkably well-established for network intrusion detection [9], [10]. However, even though the malicious purpose of a cyber-attack causes it to have distinct characteristics that could be recognized in a thorough analysis by security practitioners, an attacker can create perturbations in IoT network traffic to deceive these algorithms and be misclassified as benign. The increasing threat posed by adversarial attacks restates the need for better defense strategies for intelligent IoT network intrusion detection systems [11], [12].

To ensure that ML is used in a secure way, organizations should proactively search for vulnerabilities in their intelligent systems. By simulating realistic attack vectors, ML engineers and security practitioners can anticipate possible threats and use that knowledge to improve their countermeasures [13]. But throughout the current scientific literature, various studies apply adversarial evasion attacks to intrusion detection and provide the examples as direct input to an ML model without questioning if they are viable for a real deployment scenario [14], which may result in misleading robustness evaluations where a model seems to be robust because it was tested against examples that it will not encounter in real IoT network traffic [15].

This work addresses the challenge of improving the robustness of tree-based algorithms and ensembles for IoT network intrusion detection. The main contributions are: (i) a description of the types of constraints required for an adversarial cyber-attack example to be realistic, (ii) a methodology for a trustworthy robustness analysis with a realistic adversarial evasion attack vector, and (iii) an analysis of several tree-based algorithms and ensembles in binary and multi-class classification scenarios, following the proposed methodology. The initial evaluation carried out in [16] was extended to include adversarial attacks performed with the Adaptative Perturbation Pattern Method (A2PM). Three supervised algorithms, Random Forest (RF), Extreme Gradient Boosting (XGB), and Light Gradient Boosting Machine (LGBM), and one unsupervised algorithm, Isolation Forest (IFOR), were evaluated using the IoT-23 and Bot-IoT datasets. In addition to regular training, the effectiveness of performing adversarial training with realistically perturbed samples was also analyzed.

The present paper is organized into multiple sections. Section 2 provides a survey of previous work on ML robustness for IoT network intrusion detection. Section 3 describes the constraints required to achieve adversarial realism and defines a methodology for a trustworthy robustness analysis. Section 4 describes the experimental evaluation performed following the proposed methodology, including the scenarios, datasets, adversarial method, models, and evaluation metrics. Section 5 presents a comparative analysis of the results obtained by each ML model in each scenario. Finally, Section 6 addresses the main conclusions and future research topics.

## 2   Related Work

In recent years, the susceptibility of tree-based algorithms to adversarial examples has been drawing attention for network intrusion detection [17], [18]. To better protect these ML models from adversarial attacks, several defense strategies have been developed. Some attempt to improve the intrinsic robustness of entire tree ensembles at once [19], [20], whereas other address each individual decision tree at a time [21], [22]. Nonetheless, the most effective and widespread defense is adversarial training because it anticipates the data variations that an ML model may encounter [23]. Augmenting a training set with examples created by an adversarial evasion attack method enables a model to learn additional characteristics that the samples of each class can exhibit, so it becomes harder for an attacker to deceive it [24].

However, performing adversarial training with unrealistic examples will make a model learn distorted characteristics that will not be exhibited by real samples during its inference phase [25]. This raises a major security concern because training with unrealistic data may not only deteriorate a model's robustness against adversarial data, because it will not learn the subtle nuances that an attacker can exploit, but it may also be significantly detrimental to a model's generalization to regular data, leading to accidental data poisoning and to the introduction of hidden backdoors that make a model even more vulnerable to attacks [26].

Since the focus of adversarial ML has been the computer vision domain, the common attack vector is to freely exploit the internal gradients of artificial neural networks to generate random data perturbations in the pixels of an image [27], which can lead to unrealistic adversarial examples in tabular data. Consequently, most state-of-the-art evasion attack methods do not support other settings nor models that do not have loss gradients [28], which severely limits their applicability to the IoT network intrusion detection domain. To adversarially train a model and improve its robustness with realistic cyber-attack examples, a defender will need to resort to methods that support the specificities of a communication network.

Even though most methods were intended to attack images, a few could be adapted to tabular data. Both the Jacobian-based Saliency Map Attack (JSMA) [29] and the OnePixel attack [30] were developed to minimize the number of modified pixels, which could be used to solely perturb a few features in a network traffic flow. Nonetheless, the perturbations are still randomly generated, so the resulting values for those few features are commonly incompatible with the remaining features of a flow [31]. On the other hand, A2PM [32] was specifically developed for communication networks, assigning an independent sequence of adaptative patterns to analyze the characteristics of each class and create realistic data perturbations that preserve the purpose of a cyber-attack. Due to its suitability for IoT network traffic, it was selected for this work.

To determine the most adequate ML models for IoT network intrusion detection, it is important to understand the results and conclusions of previous performance evaluations. A comprehensive survey [33] analyzed studies published until 2018, highlighting the advantages and limitations of each model. Tree-based algorithms and ensembles obtained good results in the reviewed performance evaluations, although their robustness was not addressed. In more recent studies, the best performances were achieved by RF in a testbed replicating an industrial plant [34], XGB with the CIDDS-001, UNSW-NB15 and NSL-KDD datasets [35], LGBM with an industrial dataset [36], and IFOR in an IoT testbed for zero-day attacks [37]. Due to their promising results, RF, XGB, LGBM and IFOR were selected for this work.

To the best of our knowledge, no previous work has analyzed the adversarial robustness of these four algorithms against realistic adversarial examples of cyber-attacks targeting IoT systems nor the effectiveness of an adversarial training approach with realistically perturbed samples.

## 3   Adversarial Realism

This section describes the types of constraints required for an adversarial cyber-attack example to achieve realism and defines a methodology for a trustworthy adversarial robustness analysis with a realistic evasion attack vector.

### 3.1   Data Constraints

In the IoT network intrusion detection domain, cyber-attacks can be identified by analyzing the characteristics of network traffic flows, which are represented in a tabular data format. The features of a flow may be required to follow specific data distributions, according to the specificities of a communication network and the utilized protocols. Furthermore, due to their distinct malicious purposes, different cyber-attacks may exhibit entirely different feature correlations. Since a data sample must represent a real traffic flow, either benign activity or a cyber-attack class, it must fulfill all the constraints of this complex tabular data domain.

To generate adversarial cyber-attack examples that could evade detection in a real IoT system, the constraints must be carefully analyzed. For instance, a key characteristic of an IoT network traffic flow is the Inter-Arrival Time (IAT), which represents the elapsed time between the arrival of two subsequent packets. Its minimum (MinIAT) and maximum (MaxIAT) values are valuable features for the detection of several cyber-attack classes, such as Denial-of-Service (DoS). A low MinIAT can indicate a short DoS that quickly overloads a server with requests, whereas a high MaxIAT can indicate a lengthy DoS that overwhelms a server by keeping long connections open [38].

When perturbing these features, validity is essential because a successful adversarial attack is not necessarily a successful cyber-attack. If MinIAT was increased to a value higher than MaxIAT, a flow could become an adversarial example that a model would misclassify as benign. However, that would be an invalid network flow that a model would never encounter in a real deployment scenario because it could not be transmitted through a communication network. Therefore, to preserve validity within the network traffic structure, a domain constraint must be enforced: MinIAT must not be higher than MaxIAT. These types of constraints, including value ranges and multiple category membership, have started being investigated in [31] to improve the feasibility of adversarial attacks for network intrusion detection.

Nonetheless, validity is not enough for an adversarial attack to be a successful cyber-attack. It is imperative to also address class coherence. Even if the previous domain constraint was fulfilled when increasing MinIAT, the resulting flow could still not be coherent with the intended purpose of a cyber-attack class. Valid adversarial examples with increased MinIATs could be misclassified as benign, but not be quick enough to overload a server in a real scenario. Consequently, those supposed adversarial examples would not actually belong to the short DoS class. Instead, they would represent just regular traffic that would not be useful for a cyber-attack, so an ML model would be correct to label them as benign. Therefore, to preserve coherence, it is necessary to also enforce a class-specific constraint: MinIAT must not be higher than the highest known value of that feature for the short DoS class. These types of constraints are based on the idea initially introduced in [32], where data perturbations were created according to the correlation between multiple features.

Even though validity and coherence have previously been investigated, sometimes with different designations, it is pertinent to address them together in a single unifying concept: adversarial realism. Hence, for an adversarial example to be realistic, it must be valid within its domain structure and coherent with the characteristics and purposes of its class, by simultaneously fulfilling all domain and class-specific constraints. Regarding cyber-attacks targeting IoT systems, realistic adversarial examples must be valid traffic capable of being transmitted through a communication network, as well as coherent cyber-attacks capable of fulfilling their intended malicious purpose.

### 3.2 Analysis Methodology

To perform a trustworthy adversarial robustness analysis of multiple ML models, it is imperative to carry out realistic evasion attack vectors that use valid and coherent examples. The proposed methodology is meant to enable a security by design approach during the development of an intelligent system, and to be regularly replicated with new data recordings to ensure that the models continue to be adversarially robust.

Considering that network intrusion detection systems are developed in a secure environment and deployed with security measures to encapsulate the utilized models, an attacker will not likely have access neither to a model's training set nor to its internal parameters. Therefore, in addition to fulfilling all domain and class-specific constraints, an adversarial attack method will have to rely solely on a model's class predictions in a black-box or grey-box setting, depending on the available system information about the model and feature set [39]. This attack vector can be simulated by solely giving an evasion attack method access to a holdout set with IoT network traffic that a model has not yet seen. The analysis can be performed in four steps:

1. Preprocess a dataset, splitting it into training and holdout sets.
2. Train and validate an ML model, using the training set.
3. Perform an evasion attack to create a model-specific adversarial holdout set, using the regular holdout set and the model's class predictions.
4. Evaluate the model's performance on the regular and adversarial holdout sets, analyzing its generalization to regular data and its robustness to adversarial data.

In addition to a regularly trained model, an adversarial training approach can be included to analyze the trade-off of performance on regular data to improve the performance on adversarial data. The complete analysis can be performed in five steps:

1. Preprocess a dataset, splitting it into training and holdout sets.
2. Create a simple data perturbation in a copy of each sample of the regular training set, creating an augmented adversarial training set with more data variations.
3. Train and validate two ML models, the first using the regular training set and the second using the adversarial training set.
4. Perform two evasion attacks to create two model-specific adversarial holdout sets, using the regular holdout set and each model's class predictions.
5. Evaluate each model's performance on the regular and adversarial holdout sets, comparing their generalization to regular data and their robustness to adversarial data.

From the comparison performed in the last step, the model with the most adversarially robust generalization can be selected for deployment. Posteriorly, if new data is recorded, this methodology can be replicated to anticipate possible threats and use that knowledge to improve the defense strategy (see Fig. 1).

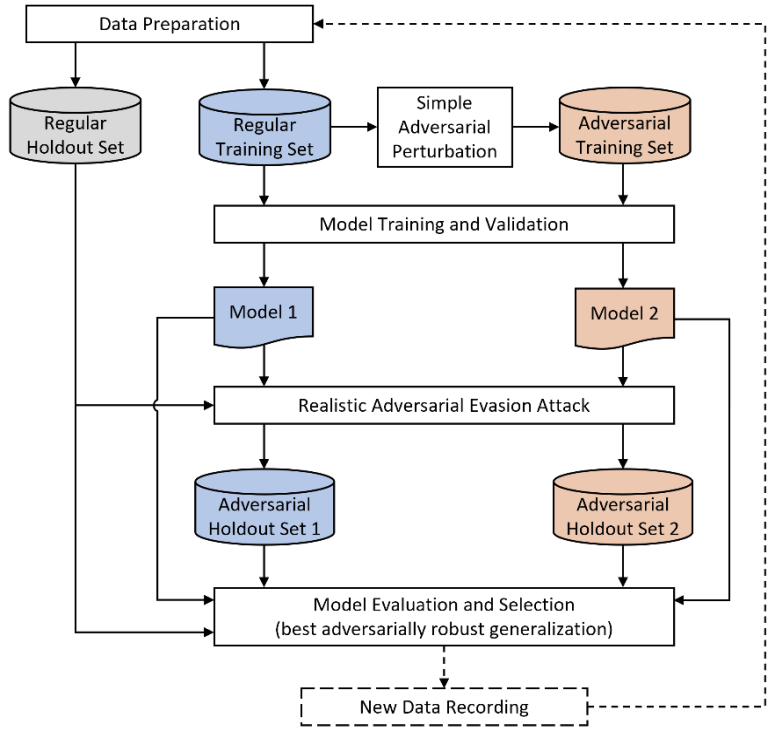

**Fig. 1.** Adversarial robustness analysis methodology.

## 4 Experimental Evaluation

This section describes the experimental evaluation performed following the proposed methodology, including the considered scenarios and datasets, and the utilized adversarial method, ML models, and performance evaluation metrics. The analysis was carried out on a machine with 16 gigabytes of random-access memory, an 8-core central processing unit, and a 6-gigabyte graphics processing unit. The implementation relied on the Python 3 programming language and the following libraries: *numpy* and *pandas* for data preparation and manipulation, *scikit-learn* for the implementation of RF and IFOR, *xgboost* for XGB, and *lightgbm* for LGBM. The previously developed *a2pm* library was used to perform a constrained adversarial example generation.

### 4.1 Scenarios and Datasets

Two distinct scenarios were considered for IoT network intrusion detection: binary and multi-class classification. In the former, the aim of a model was to detect that a network traffic flow was malicious, whereas in the latter, a model had to correctly identify each cyber-attack class and distinguish between them.

Both scenarios included the IoT-23 [40] and Bot-IoT [41] datasets. These are public datasets that contain multiple labeled captures of benign and malicious network flows within IoT systems. The recorded data is extremely valuable because it manifests real IoT network traffic patterns and includes various classes of common cyber-attacks. The former was created in the Stratosphere Research Laboratory and contains twenty-three labeled captures of malware attacks targeting real IoT devices between 2018 and 2019. Despite the latter also incorporating simulated devices and services, it resulted from a realistic testbed environment of botnet activity developed at the University of New South Wales. Table 1 provides an overview of the main characteristics of the datasets. The class labels were either Benign or the name of a cyber-attack class, such as Distributed DoS (DDoS) and Command and Control (C&C).

Table 1. Main characteristics of utilized datasets.

| Dataset | Selected Captures | Total Samples | Class Samples | Class Label |
|---|---|---|---|---|
| IoT-23 | 1-1<br>34-1 | 1,031,893 | 539,587 | POAHPS |
|  |  |  | 471,198 | Benign |
|  |  |  | 14,394 | DDoS |
|  |  |  | 6,714 | C&C |
| Bot-IoT | Full5pc-4 | 668,522 | 576,884 | DDoS |
|  |  |  | 91,082 | Recon |
|  |  |  | 477 | Benign |
|  |  |  | 79 | Theft |

A preprocessing stage was applied to both datasets, considering their distinct characteristics. First, the features that did not provide valuable information about a flow's benign or malicious purpose, such as origin and destination addresses, were discarded.

Then, one-hot encoding was employed to convert the categorical features to numeric values. Due to their high cardinality, low frequency categories were aggregated into a single category to avoid encoding qualitative values that had a small relevance. Finally, the data was randomly split into training and holdout sets with 70% and 30% of the samples, respectively. To preserve the imbalanced class proportions, the split was performed with stratification. The resulting IoT-23 sets were comprised of four classes and 42 features, 8 numerical and 34 categorical. Similarly, the Bot-IoT sets contained four classes and 35 features, 15 numerical and 20 categorical.

### 4.2  Adversarial Method

The realistic data perturbations required for a trustworthy analysis were created with A2PM [32]. It relies on sequences of adaptative patterns that learn the characteristics of each class. The patterns record the value intervals of individual features and value combinations of multiple features of tabular data. The learnt characteristics are then used to generate constrained adversarial examples that are coherent with the characteristics of their class and simultaneously remain valid within a domain.

Considering that the benign class represents regular IoT network traffic that is not part of an attack, A2PM was applied solely to samples of cyber-attack classes. The method was configured to use independent patterns for specific feature subsets, accounting for the constraints of numerical features and the correlation between encoded categorical features like the destination port, the communication protocol, and the connection flags. Then, two different functionalities were used to perform a simple perturbation and a full evasion attack. These exhibit distinct behaviors and were adapted to different data to prevent any bias in the evaluation of the adversarially trained model.

**Simple Adversarial Perturbation.** The method was adapted solely to the characteristics of the regular training set and then a single perturbation was created in a copy of each malicious sample of that set. This resulted in an adversarial training set with twice as many malicious samples as the regular set, so a model could learn not only from a recorded cyber-attack, but also from a simple variation of it.

A security practitioner could perform these simple perturbations manually by analyzing the entire dataset and adding modified samples according to the characteristics of each cyber-attack class. Nonetheless, the automated process of A2PM was preferred. When compared to the training time of an ML model, the few additional seconds required to create the simple sample variations were negligible.

**Realistic Adversarial Evasion Attack.** The method was adapted solely to the characteristics of the regular holdout set and then a full evasion attack was performed, creating as many data perturbations as necessary in a copy of each malicious sample of that set until every flow was misclassified or a maximum of 30 misclassification attempts were performed. This resulted in an adversarial holdout set with the same size as the regular set, but where each malicious sample was replaced with an adversarial example.

In the multi-class scenario, the performed adversarial evasion attacks could be untargeted, causing any misclassification of malicious samples to different classes, as well as targeted, seeking to misclassify malicious samples as the benign class. In turn, in the binary scenario, both types of evasion attacks were equivalent because all cyber-attacks were aggregated into a single class.

### 4.3    Models and Fine-tuning

The RF, XGB, LGBM, and IFOR algorithms were used to create distinct models for each dataset and scenario, which were fine-tuned through a grid search of well-established hyperparameter combinations for cyber-attack classification. To determine the optimal configuration for each model, a 5-fold cross-validation was performed. Therefore, in each iteration, a model was trained with 4/5 of a training set and validated with the remaining 1/5. The macro-averaged F1-Score was selected as the validation metric to be maximized in both regular and adversarial training, which will be detailed in the next subsection. After being fine-tuned, each model was retrained with a complete training set and evaluated using the corresponding holdout set.

**Random Forest.** RF [42] is a supervised ensemble of decision trees, which are decision support tools that use a tree-like structure. Each individual tree performs a prediction according to a specific feature subset, and the most voted class is chosen. It is based on the wisdom of the crowd, the concept that the collective decisions of multiple classifiers will be better than the decisions of just one.

The default Gini Impurity criterion was used to measure the quality of the possible node splits, and the maximum number of features selected to build a tree was the square root of the total number of features of each dataset. The optimized value for the maximum depth of a tree was 16, and the minimum number of samples required to create a leaf node was 2 and 4 for the binary and multi-class scenarios, respectively. Table 2 summarizes the configuration.

Table 2. Summary of RF configuration.

| Parameter | Value |
|---|---|
| Criterion | Gini Impurity |
| No. of estimators | 100 |
| Max. depth of a tree | 16 |
| Max. features | $\sqrt{\text{No. of features}}$ |
| Min. samples in a leaf | 2 to 4 |

**Extreme Gradient Boosting.** XGB [43] performs gradient boosting using a supervised ensemble of decision trees. A level-wise growth strategy is employed to split nodes level by level, seeking to minimize a loss function during its training.

The acknowledged Cross-Entropy loss was used for both binary and multi-class scenarios, and the Histogram method was selected because it computes fast histogram-based approximations to choose the best splits. The key parameter of this model is the learning rate, which controls how quickly the model adapts its weights to the training data. It was optimized to relatively small values for each training set and scenario, ranging from 0.01 to 0.2. Table 3 summarizes the configuration.

Table 3. Summary of XGB configuration.

| Parameter | Value |
|---|---|
| Method | Histogram |
| Loss function (objective) | Cross-Entropy |
| No. of estimators | 80 to 120 |
| Learning rate | 0.01 to 0.2 |
| Max. depth of a tree | 8 |
| Min. loss reduction (gamma) | 0.01 |
| Feature subsample | 0.7 to 0.8 |

**Light Gradient Boosting Machine.** LGBM [44] also utilizes a supervised ensemble of decision trees to perform gradient boosting. Unlike XGB, a leaf-wise strategy is employed, following a best-first approach. Hence, the leaf with the maximum loss reduction is directly split in any level.

The key advantage of this model is its ability to use Gradient-based One-Side Sampling (GOSS) to build the decision trees, which is computationally lighter than previous methods and therefore provides a faster training process. The Cross-Entropy loss was also used, and the minimum samples required to create a leaf was optimized to 16. To avoid fast convergences to suboptimal solutions, the learning rate was also kept at small values for the distinct datasets and scenarios. Table 4 summarizes the configuration.

Table 4. Summary of LGBM configuration.

| Parameter | Value |
|---|---|
| Method | GOSS |
| Loss function (objective) | Cross-Entropy |
| No. of estimators | 80 to 120 |
| Learning rate | 0.01 to 0.2 |
| Max. depth of a tree | 16 |
| Max. leaves in a tree | 32 |
| Min. loss reduction (gamma) | 0.01 |
| Min. samples in a leaf | 16 |
| Feature subsample | 0.7 to 0.8 |

**Isolation Forest.** IFOR [45] isolates anomalies through an unsupervised ensemble of decision trees. The samples are repeatedly split by random values of random features until outliers are segregated from normal observations. Unlike the previous algorithms, IFOR can only perform anomaly detection with unlabeled data. Nonetheless, it can be compared to the remaining models in the binary scenario, so cross-validation was also utilized to optimize its configuration.

This model relies on the contamination ratio of a training set, which must not exceed 50%. Hence, the number of samples intended to be anomalies must be lower than the number of remaining samples, otherwise outliers cannot be detected. To reduce the contamination of the training data, each cyber-attack class was randomly subsampled with stratification. The optimized ratios of the total proportion of malicious samples were 0.4 and 0.5 for IoT-23 and Bot-IoT, respectively. Therefore, the training data contained 40% and 50% of anomalies. Table 5 summarizes the configuration.

Table 5. Summary of IFOR configuration.

| Parameter | Value |
| --- | --- |
| No. of estimators | 100 |
| Contamination | 0.4 to 0.5 |
| Max. features | 0.9 |
| Max. samples | 256 |

### 4.4 Evaluation Metrics

To analyze a model's robustness, its performance on the regular holdout set was compared to its performance on its respective adversarial holdout set. The considered evaluation metrics and their interpretation are briefly described below [46], [47].

Accuracy is a standard metric for classification tasks that measures the proportion of correctly classified samples. It uses the True Positives (TP), True Negatives (TN), False Positives (FP) and False Negatives (FN) reported by the confusion matrix. However, its bias towards the majority classes must not be disregarded when the minority classes are particularly relevant, which is the case of cyber-attack classification. Since A2PM generated adversarial examples solely for malicious samples, even if all examples evaded detection, an accuracy as high as the proportion of benign flows could still be achieved. Therefore, to correctly exhibit the misclassifications caused by the performed attacks, the accuracy score was calculated using the samples of all classes except the benign class. It can be expressed as:

$$Accuracy = \frac{TP + TN}{TP + TN + FP + FN} \quad (1)$$

Despite the reliability of accuracy, there are other suitable metrics. For instance, precision measures the proportion of predicted attacks that were actual attacks, which indicates the relevance of a model's predictions. On the other hand, recall, which corresponds to TPR, measures the proportion of actual attacks that were correctly predicted, reflecting a model's ability to identify malicious flows. Another valuable metric is the false positive rate because it measures the proportion of benign flows that were incorrectly predicted to be attacks, leading to false alarms.

These metrics are indirectly consolidated in the F1-Score, which calculates the harmonic mean of precision and recall. A high F1-Score indicates that malicious flows are being correctly identified and there are low false alarms. It can be macro-averaged to give all classes the same relevance, which is well suited for imbalanced training data. Due to the consolidation of multiple metrics, the macro-averaged F1-Score was the preferred metric for the model fine-tuning. It is mathematically defined as:

$$Macro\text{-}averaged\ F1\text{-}Score = \frac{1}{C} * \sum_{i=1}^{C} \frac{2 * P_i * R_i}{P_i + R_i} \qquad (2)$$

where $P_i$ and $R_i$ are the precision and recall of class $i$, and $C$ is the number of classes.

## 5      Results and Discussion

This section presents the results obtained by the four tree-based algorithms in the binary and multi-class scenarios, as well as a comparative analysis of their robustness against adversarial network flow examples, with regular and adversarial training approaches.

### 5.1    Binary Classification

In the binary scenario, the models created with regular training exhibited reasonable performance declines on the IoT-23 dataset. Even though all four models achieved over 99% accuracy on the original holdout set, numerous misclassifications were caused by the adversarial attacks. The lowest score on an adversarial set, 68.35%, was obtained by XGB. In contrast, the models created with adversarial training kept significantly higher scores. By training with one realistically generated example per malicious flow, all models successfully learnt to detect most cyber-attack variations. IFOR stood out for preserving the 99.98% accuracy it obtained on the original holdout set throughout the entire attack, which highlighted its excellent generalization (see Fig. 2).

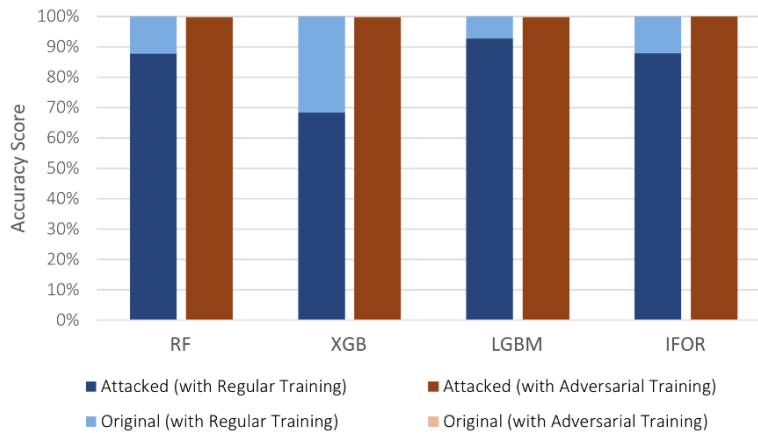

**Fig. 2.** Accuracy on IoT-23 binary classification.

Regarding the Bot-IoT dataset, the declines were significantly higher. The inability of these tree-based algorithms to distinguish between the different classes evidenced their high susceptibility to adversarial examples. The score of LGBM dropped to 26.04%, followed by IFOR, with 34.31%. Regarding the latter, it could not reach 85% in the original holdout set, possibly due to the occurrence of overfitting. Despite some examples still deceiving them, the models created with adversarial training were able to learn the subtle nuances between each cyber-attack class, which mitigated the impact of the generated examples. Apart from IFOR, the remaining models consistently achieved scores over 97%, which indicated a good robustness (see Fig. 3).

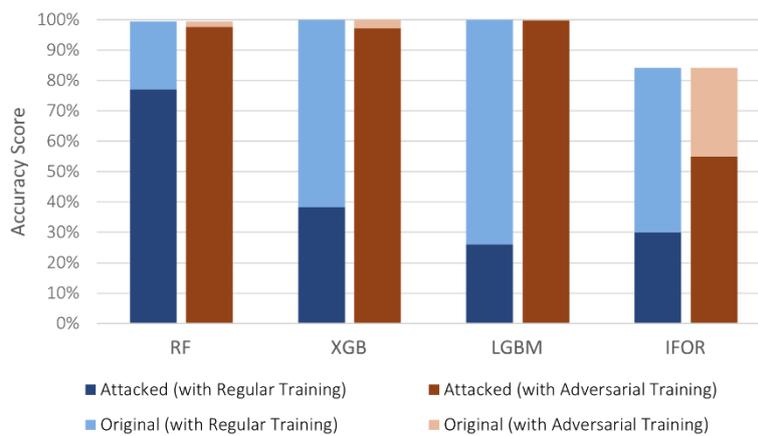

**Fig. 3.** Accuracy on Bot-IoT binary classification.

### 5.2 Multi-class Classification

In the multi-class scenario, the targeted and untargeted attacks had different impacts on a model's performance. The former caused malicious flows to be solely predicted as the benign class, whereas the latter caused malicious flows to be predicted as different classes, including other cyber-attack classes. Both attacks decreased the accuracy of the three supervised models on IoT-23, with LGBM being significantly more affected. Nonetheless, it can be observed that its targeted accuracy, 57.78%, was significantly higher than the untargeted, 32.11%, with more misclassifications occurring between different cyber-attack classes. Therefore, despite LGBM being susceptible, the benign class was more difficult to reach in multi-class intrusion detection. Even though performing adversarial training further increased the high scores of XGB, it was surpassed by RF on the targeted attack, which achieved 99.97% (see Figs. 4 and 5).

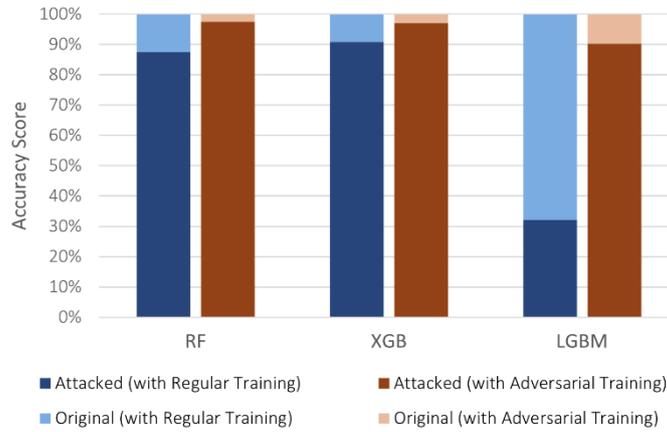

**Fig. 4.** Untargeted accuracy on IoT-23 multi-class classification.

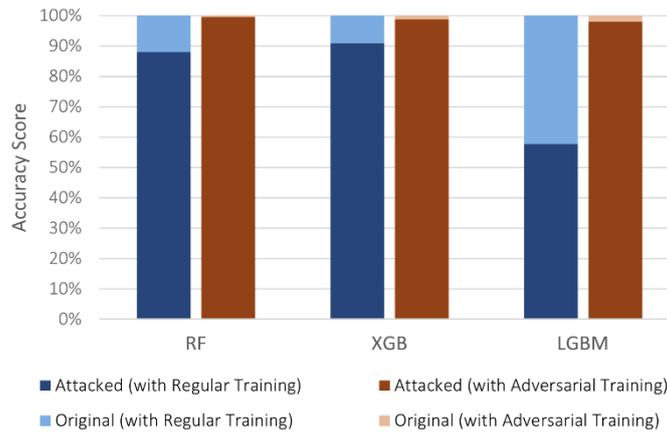

**Fig. 5.** Targeted accuracy on IoT-23 multi-class classification.

As in the previous scenario, higher declines were exhibited for the Bot-IoT dataset. The untargeted attacks performed by A2PM dropped the accuracy of RF and XGB to near 65%, although the targeted attacks only decreased it to 87.50% and 97.14%. Adversarial training contributed to the creation of more robust models, leading to fewer incorrect class predictions. Regarding RF, it could even preserve the 99.98% score it obtained on the holdout set throughout the entire attack. Even though some malicious flows still evaded detection, the robustness of both XGB and LGBM was also successfully improved. Overall, the adversarial robustness of the analyzed tree-based algorithms was significantly improved by augmenting their training data with a simple variation of each cyber-attack (see Figs. 6 and 7).

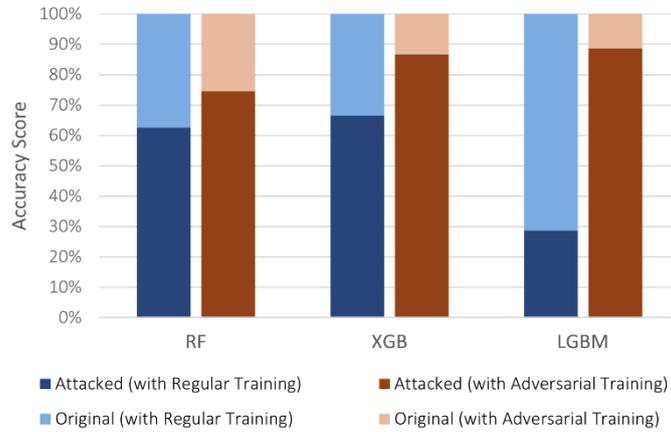

**Fig. 6.** Untargeted accuracy on Bot-IoT multi-class classification.

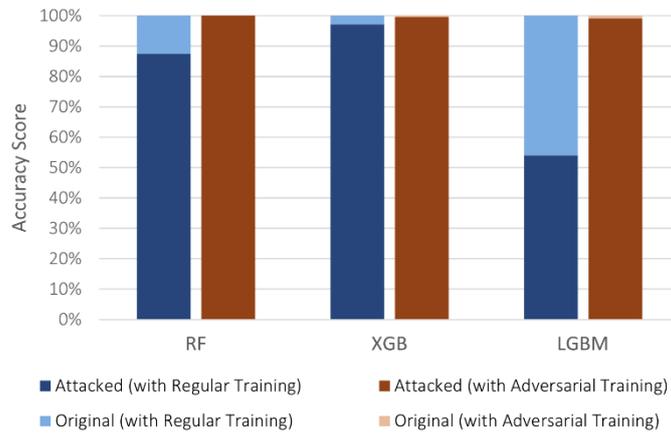

**Fig. 7.** Targeted accuracy on Bot-IoT multi-class classification.

## 6   Conclusions

This work addressed the use of ML for IoT network intrusion detection from an adversarial robustness perspective. The types of constraints required for an adversarial cyber-attack example to be valid and coherent were described, and a methodology was proposed for a trustworthy adversarial robustness analysis. The methodology was followed to analyze the robustness of four algorithms, RF, XGB, LGBM, and IFOR, using the IoT-23 and Bot-IoT datasets. Targeted and untargeted adversarial evasion attacks were performed with A2PM, and both regular and adversarial training approaches were evaluated in binary and multi-class classification scenarios.

The models created with regular training exhibited significant performance declines, which were more prominent on the Bot-IoT dataset. Even though RF was the least affected in the binary scenario, XGB consistently achieved the highest accuracy on multi-class classification. Furthermore, when adversarial training was performed, all four models successfully learnt to detect most cyber-attack variations and kept significantly higher scores when attacked. The adversarially trained IFOR and RF stood out for preserving the highest accuracy throughout entire attacks, on binary IoT-23 and multi-class Bot-IoT, respectively. Regarding LGBM, the obtained results suggest that it is highly susceptible to adversarial examples, especially on imbalanced multi-class classification. Nonetheless, this vulnerability can be successfully tackled by augmenting its training data with one realistic adversarial example per malicious flow.

The performed analysis evidenced the inherent susceptibility of tree-based algorithms to adversarial examples and demonstrated that they can benefit from defense strategies like adversarial training to create more robust models. In the future, it is pertinent to further contribute to robustness research by replicating this methodical analysis with novel datasets, ML models, and evasion attack methods. As the threat of adversarial attacks increases, defense strategies must be improved and a security by design approach must be followed to ensure that ML models can provide a reliable and robust IoT network intrusion detection and cyber-attack classification.

**Author Contributions.** Conceptualization, J.V. and I.P.; methodology, J.V.; software, J.V.; validation, E.M. and I.P.; investigation, J.V. and E.M.; writing, J.V. and E.M.; supervision, I.P.; project administration, I.P.; funding acquisition, I.P. All authors have read and agreed to the published version of the manuscript.

**Funding.** The present work was partially supported by the Norte Portugal Regional Operational Programme (NORTE 2020), under the PORTUGAL 2020 Partnership Agreement, through the European Regional Development Fund (ERDF), within project "Cybers SeC IP" (NORTE-01-0145-FEDER-000044). This work has also received funding from UIDB/00760/2020.

**Data Availability.** Publicly available datasets were analyzed in this work. The data can be found at: IoT-23 (https://www.stratosphereips.org/datasets-iot23), Bot-IoT (https://research.unsw.edu.au/projects/bot-iot-dataset). A publicly available method was utilized in this work. The method can be found at: A2PM (https://github.com/vitorinojoao/a2pm).

**Conflicts of Interest.** The authors declare no conflict of interest. The funders had no role in the design of the study; in the collection, analyses, or interpretation of data; in the writing of the manuscript, or in the decision to publish the results.